\documentclass[preprint,showpacs,preprintnumbers,amsmath,amssymb]{revtex4}

\usepackage{graphicx}
\usepackage{dcolumn}
\usepackage{bm}

\begin{document}

\title{Phase transitions and superuniversality in the dynamics of a self-driven particle}

\author{R. Grima $^{1,2}$}

\affiliation{$^{1}$ School of Informatics, Indiana University,
Bloomington, IN 47406  \\ $^{2}$ Biocomplexity Institute, Department
of Physics, Indiana University, Bloomington, IN 47405}

\begin{abstract}
We study an active random walker model in which
a particle's motion is determined by a self-generated
field. The field encodes information about the particle's
path history. This leads to either self-attractive or self-repelling
behavior. For self-repelling behavior, we find a
phase transition in the dynamics: when the coupling between the
field and the walker exceeds a critical value, the particle's behavior changes from
renormalized diffusion to one characterized by a diverging diffusion
coefficient. The dynamical behavior for all cases is surprisingly independent
of dimension and of the noise amplitude.
\end{abstract}


\maketitle

Active walkers are random walkers whose motion is determined by a
potential surface which is amenable to change by the walkers
themselves \cite{Schweitzer1}. They are one of several different
classes of random walks in which walkers interact with their path
history. The memory of these walkers leads to interesting
properties, uncommon to pure random walks. Other common types of
such self-interacting walks are the reinforced random walk
\cite{Davis,Pemantle} and the self-avoiding random walk
\cite{Amit,Schulz}.

Models based on active walkers have been used to study a number of
phenomena, including ant foraging patterns \cite{Schweitzer2},
traffic \cite{Helbing2}, the formation of human and animal trail
systems \cite{Helbing1}, animal mobility \cite{Ebeling}, chemotactic
aggregation \cite{Grima,Newman} and the self-assembly of networks
\cite{Schweitzer3}. The advantage of formulating problems in such a
formalism is that it enables analytic study using methods from
statistical physics and the general theory of stochastic processes.
In this article, we analyze in detail a fairly generic type of
active walker model, in which a particle interacts with its path
history by means of a self-generated field. This model is related to
the self-driven many particle system first introduced by Schweitzer
and Schimansky-Geier \cite{Schweitzer1}, which studied it by means
of a mean-field approximation applied on the stochastic model.
Newman and Grima \cite{Newman} analyzed the effect of fluctuations
on the system's dynamics by means of a many-body theory approach.
The spatio-temporal correlations make the problem difficult to
understand, usually confining analysis to the case of weak coupling
between particles and the generated field \cite{Newman} or to the
case of small noise \cite{Grima}. Indeed even the case of a single
self-driven particle is non-trivial \cite[see for
example]{Grima,Schulz,Tsori}. We study a variant of the single
self-driven particle model and develop a general theory to elucidate
the particle's rich and complex dynamical behavior. This approach
differs from the previous ones, in that it is valid for all coupling
strengths and for both weak and strong noise. The model is
potentially applicable to understanding a chemotactic biological
system under certain conditions, a topic briefly discussed at the
end of this article.

Consider a random walker whose motion is described by the following
coupled equations:
\begin{align}
\label{e1}
m {\ddot {\bf x}_{c}}(t)+ {\dot {\bf x}_{c}}(t) &= {\bf \xi}(t) + \kappa \alpha \nabla \ln \phi({\bf x}_{c},t) , \\
\label{e2}
\partial _{t} \phi({\bf x},t) &= D_{1}\nabla ^{2} \phi({\bf x},t) - \lambda \phi({\bf x},t)
+\beta \delta ({\bf x}-{\bf x}_{c}(t)).
\end{align}
Eq.(\ref{e1}) is a Langevin equation describing the motion of a
walker with mass $m$. The stochastic variable $\xi$ is white noise
defined through the statistical averages: $\langle \xi^{i}(t)
\rangle = 0$ and $\langle \xi^{i}(t)\xi^{j}(t') \rangle = 2 D_0
\delta_{i,j} \delta(t-t')$, where $i$ and $j$ refer to the spatial
components of the noise vectors. The mass is assumed to satisfy the
condition $m \ll 1$, implying that in the absence of
self-interaction the walker's dynamics are of the over-damped type.
The walker's diffusion coefficient in this case is $D_0$.
The self-generated field is denoted by $\phi$; its temporal dynamics
are determined by the reaction-diffusion equation Eq.(\ref{e2}). We shall
refer to the field as a chemical field since this is most consistent
with a physical interpretation of Eq.(\ref{e2}). Then the latter equation
describes the continuous
local release of chemical by the walker at a rate $\beta$, the
diffusion of the chemical with diffusion coefficient $D_1$ and its
decay at a constant rate $\lambda$.

The self-interaction comes from the second term on the R.H.S. of
Eq.(\ref{e1}). This term implies that the walker's motion is partly
determined by the local gradient of the field. Note that the field
encodes information about the walker's path history, meaning that
the walker's motion at any given time is a complicated function of
its previous whereabouts. The strength of the coupling between the
field and the walker's motion is determined by $\alpha$. The
constant $\kappa$ can take values $1$ or $-1$: for $\kappa = 1$, the
walker tends to explore regions already visited (a self-attracting
walker) whereas for $\kappa = -1$ the opposite is true (a
self-repelling walker). We assume that the initial chemical
concentration is described by some function $\phi_0({\bf x},t)$
which is greater than zero at all points in space.

The Langevin formulation is not usually considered the most
convenient representation for the purposes of analytic calculations
and so it is customary to derive a differential equation for the
single-particle probability distribution \cite [for example] {Newman}.
For the problem at hand, this approach does not permit much analytic
progress in understanding the walker's behavior. Applying a mean
field approximation on the differential equation for the single-particle
probability distribution one obtains a Fokker-Planck type equation,
which then permits a perturbative analysis in a coupling parameter.
This approach ignores the important spatio-temporal correlations
inherent in the problem and enables one to understand the walker's
behavior only when the coupling strength $\alpha$ is
very small. The non-markovian nature of the problem makes its
solution a challenging task.

We here present a simple method to extract the asymptotic behavior
of the active walker model. The results can also be reproduced by
the method described in \cite{Grima}. However the method to be
presented here is more transparent and gives a physically tractable
picture of the complex underlying dynamics. Its main advantages are
that spatio-temporal correlations are not ignored and that it
enables an understanding of the motile behavior for all values of
the coupling strength and of the other parameters. We
start by switching to a description in discrete time $t = n \Delta
t$ where $n \in \Bbb{N} $. Space is continuous. The walker's
position at time $t+\Delta t$ is determined by the gradient of the
logarithm of the chemical concentration it measures at time $t$.
Since the walker secretes an amount of chemical $\beta \Delta t$ at
every time step, then if the walker is at position ${\bf x}_c (t)$
at time $t$, the chemical field sensed by the walker at time $t$ is
given by:
\begin{equation}
\label{sum_phi} \phi = \sum_{n=1}^{t/\Delta t} \frac{\beta \Delta
t}{(4 \pi D_1 n \Delta t)^{d/2}} \exp \bigg[-\lambda n \Delta t -
\frac{\sum_{i=1}^{d} [x_c^i(t)-x_c^i(t-n \Delta t)]^2}{4 D_1 n
\Delta t} \biggr] = \sum_{n=1}^{t/\Delta t} \phi_n,
\end{equation}
and the gradient of the field is given by:
\begin{equation}
\label{sum_phigrad} \nabla \phi = - \sum_{n=1}^{t/\Delta t}
\frac{[{\bf x}_c(t)-{\bf x}_c(t-n \Delta t)]}{2 D_1 n \Delta t}
\phi_n,
\end{equation}
where $x_c^i(t)$ is the $i^{th}$ component of the particle position
vector ${\bf x}_c(t)$ and $d$ is the dimensionality of the space in
which particle movement occurs. Since the chemical decays in a time of
the order $1/\lambda$ then the concentration at time $t$ will be
approximately determined by the previous positions of the walker at
times $t' > t - 1/\lambda$. This implies that the sum in
Eq.(\ref{sum_phi}) and Eq.(\ref{sum_phigrad}) can be truncated at
$n_{max} = 1/\lambda \Delta t$. Now consider the term $x_c(t-n
\Delta t)$. Since we are interested in the walker's behavior in the
asymptotic limit $t \gg 1/\lambda$, then $n \Delta t \le n_{max}
\Delta t = 1/\lambda \ll t$. Thus it is possible to replace the term
$x_c^{i}(t-n \Delta t)$ in the above two equations by its Taylor
series expansion. Keeping terms only to first order in $\Delta t$ we
have:
\begin{equation}
\nabla \ln \phi({\bf x}_{c},t) = \frac{\nabla \phi({\bf
x}_{c},t)}{\phi({\bf x}_{c},t)} = -\frac{\dot {\bf x}_c(t)}{2D_1}.
\end{equation}
Thus from Eq.(\ref{e1}) and the above equation, it follows that for
long times, the behavior of the walker is dominated by the effective
Langevin equation:
\begin{equation}
\label{mod_Lan} m {\ddot {\bf x}_{c}}(t)+ \biggl(1 + \frac{\kappa
\alpha}{2 D_1} \biggr) {\dot {\bf x}_{c}}(t) = {\bf \xi}(t).
\end{equation}
It is not possible to systematically calculate corrections to this
equation by keeping more terms in the Taylor expansion of the
position terms. It can however be shown that such corrections are
negligible in high dimensions, $d \gg 2$. These issues are discussed
more fully in the Appendix.

The modified Langevin equation Eq.(\ref{mod_Lan}) is clearly valid
after some time $t^*$ such that $t^* \gg 1/\lambda$. We define
$\gamma = (1 + \kappa \alpha / 2 D_1)/m$ and integrate
Eq.(\ref{mod_Lan}) to get an equation for the time evolution of the
$i^{th}$ component of the velocity vector:
\begin{equation}
{\dot {x}_{c}^{i}}(t) = {\dot {x}_{c}^{i}}(t^*) \exp{[-\gamma
(t-t^*)]} + \frac{1}{m} \int_{t^*}^{t} dt' \exp{[-\gamma(t-t')]}
\xi^i(t').
\end{equation}
Then it follows that the velocity auto-correlation function is given
by:
\begin{align}
\nonumber \langle {\dot {x}_{c}^{i}}(t) {\dot {x}_{c}^{i}}(s)
\rangle = {\dot {x}_{c}^{i}}(t^*)^{2} &\exp{[-\gamma (t+s-2t^*)]} \\
&+ \frac{D_0}{\gamma m^2} \biggl [ \exp{(-\gamma|s-t|)} -
\exp{[-\gamma(s+t-2t^*)]} \biggr].
\end{align}
Using a Green-Kubo relation $D_R = \int_{t^*}^{t} dt' \langle {\dot
{x}_{c}^{i}}(t) {\dot {x}_{c}^{i}}(t') \rangle $ it is possible to
determine the effective (renormalized) particle diffusion coefficient
$D_R$:
\begin{align}
\nonumber D_R = \frac{{\dot {x}_{c}^{i}}(t^*)^{2}}{\gamma}
&\exp{[-\gamma (t-t^*)]}(1 - \exp{[-\gamma (t-t^*)])} \\ &+
\frac{D_0}{\gamma^2 m^2} \biggl [1 + \exp{[-2\gamma (t-t^*)]} - 2
\exp{[-\gamma (t-t^*)]} \biggr],
\end{align}
which evaluated in the limit ${t \rightarrow \infty}$ leads us to
the final set of results:
\begin{equation}
\begin{array}{rclcrcl}
\kappa &=& 1  & \forall \alpha   & D_R &=& D_0 \bigl(1 + \frac{\alpha} {2 D_{1}} \bigr)^{-2} \\
\kappa &=& -1 & \alpha < 2 D_{1} & D_R &=& D_0 \bigl(1 - \frac{\alpha} {2 D_{1}} \bigr)^{-2} \\
\kappa &=& -1 & \alpha > 2 D_{1} & D_R &=& \infty
\end{array}
\end{equation}
We defer a discussion of the physics behind these results for later.
For the moment we focus on the numerical validation of the theory.

For the case of a self-attracting walker ($\kappa = 1$) it is
predicted that: (i) the asymptotic behavior is diffusion with a
renormalized diffusion coefficient whose magnitude decreases with
increasing values of the coupling parameter $\alpha$, (ii) that the
behavior is independent of dimension (superuniversality in the velocity
correlations). We test
these predictions by numerically integrating the model equations
Eq.(\ref{e1}) and Eq.(\ref{e2}) (see Fig.1). The diffusion
coefficient in all simulations is calculated from the slope of plots
of the variance versus time for the time range $t \in (10,1000)$.
The initial chemical concentration is a Gaussian centered at the
origin, though any non-zero function is suitable.

\begin{figure} [h]
\includegraphics [width=5in]{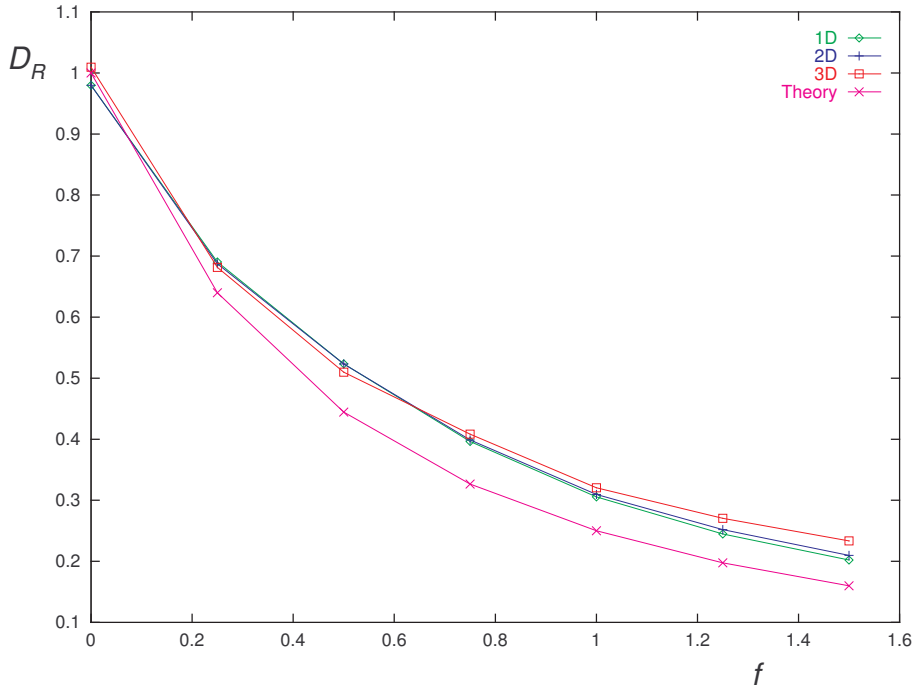}
\caption{(Color online) Plot of the renormalized diffusion
coefficient $D_R$ versus the non-dimensional coupling parameter $f$,
where $f = \alpha / 2 D_1$. The parameter $\kappa$ equals one,
implying that the walker has a tendency to explore previously
visited spatial regions. The other parameters are $m = 10^{-6}$,
$D_0=1, D_1=1, \lambda=1$, $\beta = 1$ and $\delta t = 0.1$ with
$10^{4}$ samples. As predicted, the renormalized diffusion
coefficient is the same in one, two and three dimensions.}
\end{figure}

As expected, we find that the asymptotic behavior is diffusion
characterized by renormalized diffusion coefficients which are
relatively independent of the dimension (Fig. 1). However there is some
discrepancy between the theoretical values of $D_R$ and the ones
obtained from the numerics. Regression of the one-dimensional data in
Fig. 1. shows that the numerical data is best fit by an equation of the form (see
Fig. 2): $D_R = D_0 (1 + k \alpha / 2 D_{1} )^{-2}$ where
$k=0.81 \pm 0.01$ ($10^4$ realizations). The difference between this value and the theoretical value
of unity, stems from a combination of the approximations
used in deriving the effective Langevin equation Eq.(\ref{mod_Lan})
and numerical error due to a finite time step (Note that the simulations
are off-lattice and thus there is no numerical error due to a finite spatial step). As discussed in the
Appendix, it is not possible to systematically calculate corrections to the effective
Langevin equation. However by repeating the simulations with a time step
an order of magnitude smaller than those in Fig. 1 and Fig. 2, we find that
the value of $k$ increases to $k = 0.87 \pm 0.02$ ($2\times 10^3$ realizations), which is closer to the
theoretical value. Hence it is probable that the discrepancies
between numerics and theory are in significant part due to numerical error rather than to
the approximations implicit in deriving Eq.(\ref{mod_Lan}). This is plausible since
the next order correction to Eq.(\ref{mod_Lan}) is proportional to the walker's
acceleration $\ddot {\bf x}_c(t)$ (see Eq.(\ref{second_correction}) in the Appendix) which is negligible
for a self-attracting walker since the dynamics are over-damped for small coupling ($m \ll 1$)
and become more strongly over-damped as the coupling increases (this is since the coefficient of the velocity
term in Eq.(\ref{mod_Lan}) increases with the coupling).

\begin{figure} [h]
\includegraphics [width=5in]{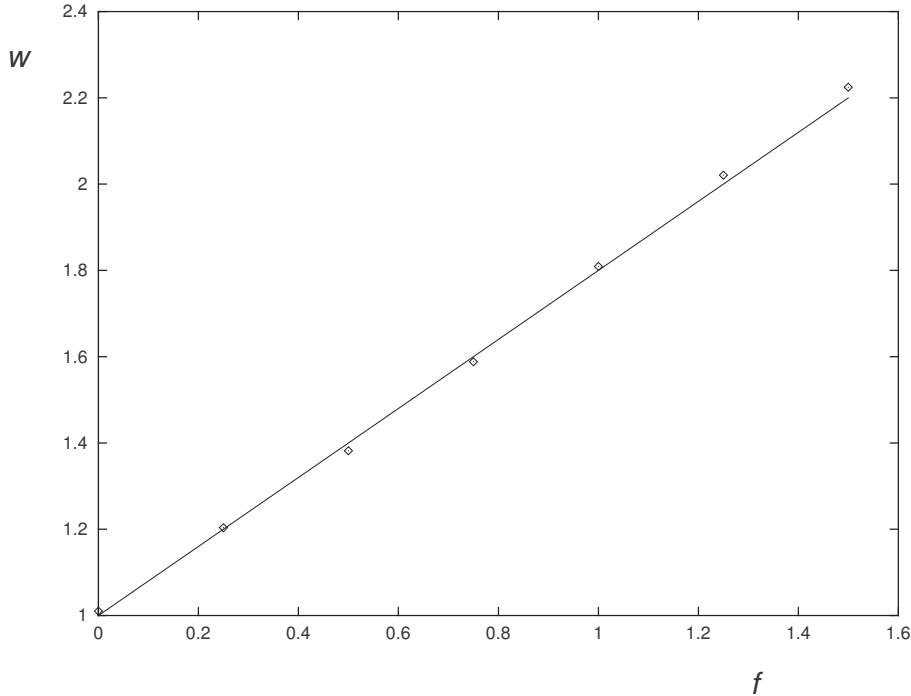}
\caption{Plot of $w = \sqrt{D_0/D_R}$ versus the non-dimensional
coupling parameter $f$, where $f = \alpha / 2 D_1$, for data
obtained from 1D simulations. Parameter values as in Fig. 1.
This verifies the functional form predicted by theory.
The solid line is the best fit through the data points. This line has a
gradient of $0.81$ and intercept of $0.99$ -- theory predicts a
gradient of $1.00$ and an intercept of $1.00$.}
\end{figure}

For the case of a self-repelling walker ($\kappa = -1$) theory
predicts that: (i) if the coupling is less than a critical
threshold, $\alpha < 2D_1$, then the asymptotic behavior is
diffusion with a renormalized diffusion coefficient, (ii) if the
coupling is above this threshold, $\alpha > 2D_1$, the particle
diffusion coefficient diverges (iii) the behavior is independent of
dimension. We tested these predictions by simulations. We find that
near the predicted singularity, $0.9 < \alpha / 2 D_1 < 1$, the
variation of $D_R$ with coupling is given by:
\begin{equation}
\label{e12} D_R/D_0 \propto {\Bigl(1 - \frac{\alpha} {2 D_{1}}
\Bigr)}^{-\zeta},
\end{equation}
where $\zeta = 2.10 \pm 0.03$ in one dimension, $\zeta = 2.23 \pm
0.02$ in two dimensions and $\zeta = 2.10 \pm 0.02$ in three
dimensions (Fig.3). These estimates agree well with the theoretical
value of $\zeta = 2$. The constant of proportionality in
Eq.(\ref{e12}) is dependent on dimension, a feature not predicted by
theory -- interestingly, as shown in Fig. 3, the data in three
dimensions is the one closest to the exact theoretical result. However
these features are not completely unexpected. This is since the coefficient of the velocity
term in Eq.(\ref{mod_Lan}) decreases with increasing coupling, meaning that the acceleration
of the walker becomes a determining factor as the critical coupling is approached. Thus the
next order correction to Eq.(\ref{mod_Lan}) (see Eq.(\ref{second_correction}) in the Appendix)
is probably not negligible (unlike the case of a self-attracting walker). As discussed in the Appendix,
these corrections are small in high dimensions and thus mostly significant in low dimensions.
These theoretical arguments support the numerical data in Fig. 3.

\begin{figure} [h]
\includegraphics [width=5in]{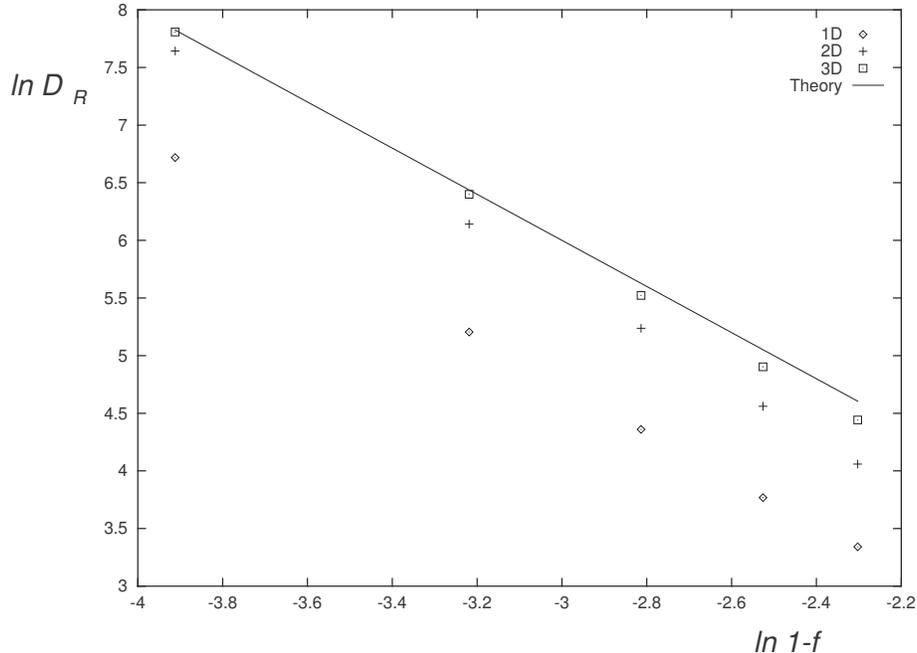}
\caption{Plot of the natural log of the renormalized diffusion
coefficient $D_R$ versus the natural log of the non-dimensional
parameter $1-f$, where $f = \alpha / 2 D_1$. Note that the $f$
values in this plot vary between 0.9 and 0.98, meaning that we are
exploring the walker's behavior near the theoretically predicted
singularity at $f = 1$. The parameter $\kappa$ equals minus one,
implying that the walker has a tendency to explore previously
unvisited spatial regions. The other parameters values are exactly
as in Fig. 1. This graph confirms that $D_R \propto (1-f)^{-2}$
implying a singularity at $f = 1$ in all dimensions.}
\end{figure}

\begin{figure} [h]
\includegraphics [width=5in]{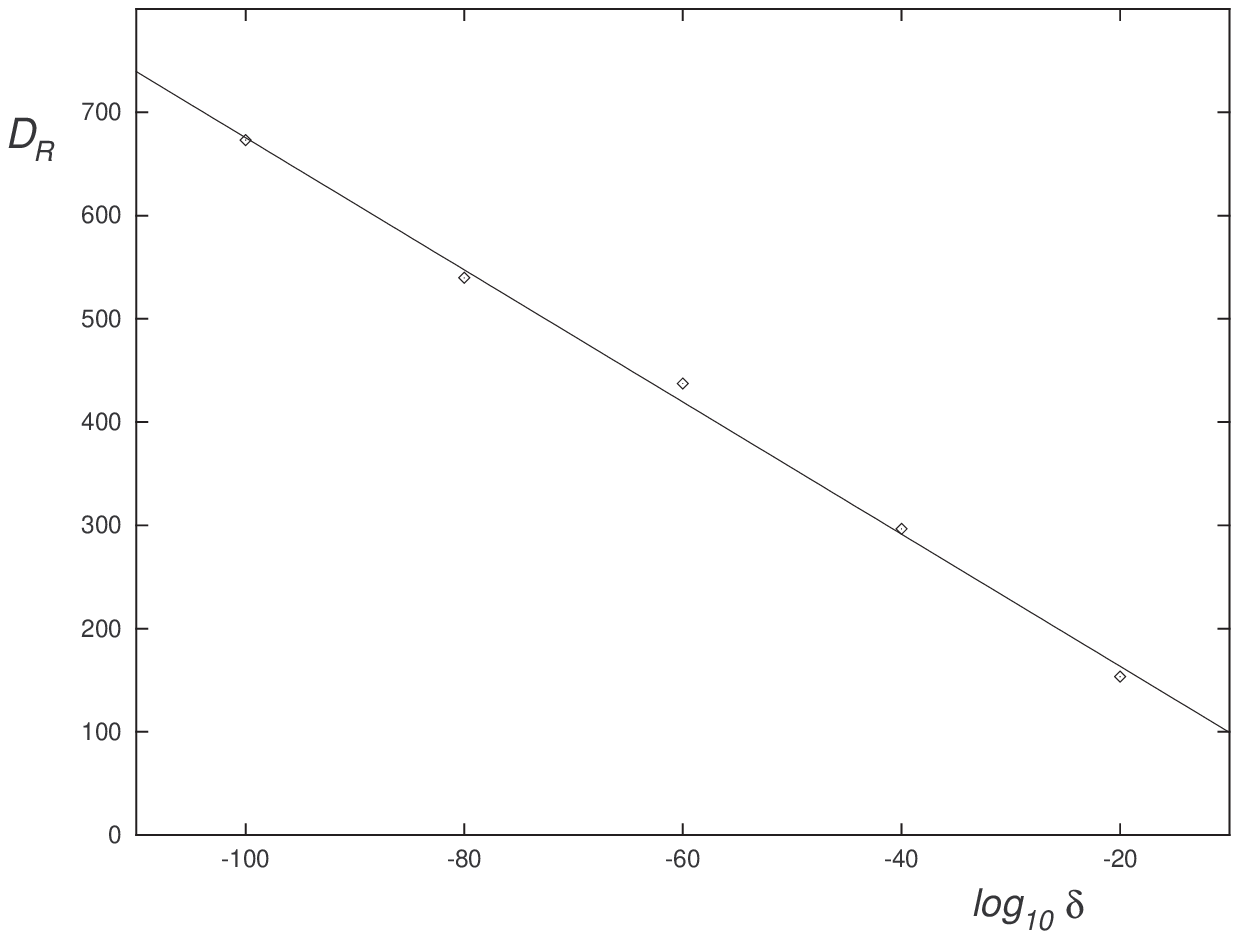}
\caption{Determination of the renormalized diffusion coefficient
$D_R$ for $f = 3$ using the same parameters as Fig. 3. The model
simulated is that given by Eq. (\ref{e13}) and Eq. (\ref{e14}). The
solid line is the best fit through the data points, indicating that
$D_R \propto \log_{10}(\delta^{-k})$ where $k$ is some positive
number. This suggests that in the limit $\delta \rightarrow 0$, $D_R
\rightarrow \infty$ for $f > 1$.}
\end{figure}

Now we numerically explore the walker's asymptotic behavior for
$\alpha / 2D_1 > 1$. In this regime, the simulations break down
after a few time steps. Using smaller values of the numerical time
step does not help much and makes the numerical analysis
computationally very expensive. This is overcome by simulating a
model given by the equations:
\begin{align}
\label{e13} {\dot {\bf x}_{c}}(t) &= {\bf \xi}(t) - \alpha \frac
{\nabla
{\phi({\bf x}_{c},t)}}{\delta + \phi({\bf x}_{c},t)}, \\
\label{e14}
\partial _{t} \phi({\bf x},t) &= D_{1}\nabla ^{2} \phi({\bf x},t) - \lambda \phi({\bf x},t)
+\beta \delta ({\bf x}-{\bf x}_{c}(t)).
\end{align}
Since we are really interested in the case $\delta = 0$, we obtained
data for several small values of $\delta$ in the hope that we could
eventually extrapolate to the desired limit. As shown in Fig.4, we
find that for $\alpha / 2 D_1 = 3$ the renormalized diffusion
coefficient clearly tends to infinity as $\delta \rightarrow 0$.
This is found to be generally true for $\alpha / 2 D_1 > 1$, meaning
that the purported transition at $\alpha / 2 D_1 = 1$ is from finite
$D_R$ to $D_R = \infty$, in agreement with theory. Note that
transitions in random walks with positive long-correlations are
known \cite{Hod}, though in this case the dynamical transition is
from normal-diffusion to super-diffusion in a one-dimensional space.

Now we discuss the theoretical results from a physical perspective.
In the absence of self-interaction, the walker's motion is
determined by the frictional force which is directly proportional to
its velocity rather than by inertia (over-damped dynamics). For the
case of a self-attracting walker ($\kappa = 1$) it is expected that
the walker spans space more slowly than for the case of no
self-interaction: this typically means sub-diffusive behavior or
diffusive behavior with a renormalized diffusion coefficient smaller
than $D_0$. We have shown that to a first approximation obtained by
truncating the walker's memory to a time of the order $1/\lambda$,
the self-interaction leads to a renormalization of the frictional
force. This implies that the dynamics are always over-damped and
that the asymptotic behavior is that of renormalized diffusion, not
sub-diffusion.

For the case of a self-repelling walker ($\kappa = -1$) it is
expected that the walker spans space faster than for the case of no
self-interaction: this typically means super-diffusive behavior or
diffusive behavior with a renormalized diffusion coefficient greater
than $D_0$. We have shown that the self-interaction leads to a
renormalization of the magnitude of the frictional force experienced
by the walker. The frictional force decreases with increasing
coupling $\alpha$ between the walker's motion and the chemical field
until at a particular value of the coupling, $\alpha = 2D_1$, the
frictional force is exactly zero and the dynamics of the walker are
purely determined by the inertial force. Thus the walker's behavior
changes from one characterized by a low-Reynolds number for weak
self-interaction to behavior characterized by a high Reynolds number
as one approaches the critical coupling. When the coupling exceeds the critical
value, we find that the walker experiences a force which is
proportional to the velocity but which has the opposite effect of
damping; the random fluctuations in the velocity due to the
stochastic force are amplified rather than suppressed and thus the
velocity of the walker increases uncontrollably with time, leading
to an infinite velocity. This is the underlying reason for the
divergence of the particle diffusion coefficient in this parameter
regime.

It is worthwhile to compare the walker's behavior with logarithmic
response to the behavior with linear response. The linear
response model was studied by Grima \cite{Grima}, who found that if the particle diffusion
coefficient (in the absence of self-interaction) is small then (i)
for a self-attracting walker, the asymptotic behavior is that of
renormalized diffusion (ii) for a self-repelling walker, the
asymptotic behavior is renormalized diffusion below a critical
threshold and ballistic diffusion above this threshold. The first
difference to be emphasized between the two models with different
responses is that for the logarithmic response the results are
generally valid for any value of the particle diffusion coefficient and
of the other parameters, whereas for the linear response model the
results are restricted to small particle to chemical diffusion
coefficients (small noise analysis). In other words the logarithmic
response gives rise to behavior which is independent of the amplitude
of the noise, an unusual property -- for example, for intermediate to
large noise, simulations indicate that the dynamics of a particle with linear response are not
that of renormalized diffusion. It is also to be noted that the value of the critical coupling in
the linear model is dimensionally dependent, unlike the superuniversal behavior in
the present case. The logarithmic response $\nabla \phi/\phi$ is weaker
than the linear response $\nabla \phi$ when $\phi > 1$ and stronger
otherwise. For the case of a self-attracting walker, the walker
tends to stay in spatial regions which it has previously visited,
meaning that the sampled chemical concentration $\phi$ at all times
is significant and not small; thus in this case we expect that
the walker with linear response to exhibit a stronger or at least
equally strong perturbation of its motion compared to that with
logarithmic response. This reasoning agrees with our results. For
the case of a self-repelling walker, the walker tends to avoid
spatial regions which it has previously visited, meaning that the
sampled chemical concentration tends to be very small; thus in this
case we expect the walker with logarithmic response to exhibit a
stronger or at least equally strong perturbation of its motion
compared to that with linear response. This qualitatively explains
the onset of the diverging diffusion regime for logarithmic response
compared to the onset of ballistic diffusion for linear response.

As we mentioned in the introduction, the model has potentially some
applications in biology. The self-driven many particle system is a
model for chemotactic aggregation or dispersion, behavior exhibited
by a number of organisms, such as the slime mould \cite{Armitage}.
The particles are then motile cells which secrete chemical and which
simultaneously move up (or down) gradients of this chemical. The
self-driven single particle can be thought of as a model for a
chemotactic cell which is away from the bulk of other cells; in that
case the cell will predominantly sense its own chemical rather than
that of other cells. In the linear response models studied by
several authors, the cell's average velocity is assumed to be
linearly proportional with the gradient of the local chemical field.
In our model, the response is logarithmic which is known to be more
realistic in some specific cases. This type of response is
frequently referred to as the Weber-Fechner law and is thought to
describe the sensory adaptation of a number of chemotactic cells to
chemotactic signals, over certain concentration ranges
\cite{Brown,Keller,Levine}. With this proviso, our present model may
perhaps be applicable to understanding some features of sensory
adaptation at the micro-organism level.

The author would like to thank Alessandro Flammini and Alessandro
Vespignani for a critical review of the manuscript and Timothy Newman for interesting discussions. This work was supported by a grant from the Faculty Research Support Program from the OVPR, Indiana University
(Bloomington Campus).

\newpage

\section*{Appendix}

It is in principle possible to calculate the next order correction
to Eq.(\ref{mod_Lan}) by retaining terms to second order in the
Taylor series expansion of $x_c^{i}(t-n \Delta t)$ in
Eq.(\ref{sum_phi}) and Eq. (\ref{sum_phigrad}). Furthermore since
the exponent is slowly varying, the the sums over $n$ can be
approximated by integrals, leading to:
\begin{equation}
\label{second_correction} \nabla \ln \phi({\bf x}_{c},t) =
-\frac{\dot {\bf x}_c(t)}{2D_1} + \frac{\ddot {\bf x}_c(t) \Delta
t}{4D_1} \left[\int_{n=1}^{n_{max}} n^{1-d/2} f(n) \bigg{/}
\int_{n=1}^{n_{max}} n^{-d/2} f(n) \right],
\end{equation}
where
\begin{equation}
\label{exp_func} f(n) = \exp \left [ - \Delta t \left(\lambda +
\sum_{i=1}^{d} \frac{\dot {x}_c^{i}(t)^2}{4 D_1} \right) n + \Delta
t^2 \left ( \sum_{i=1}^{d} \frac{\dot { x}_c^{i}(t) \ddot
{x}_c^{i}(t)}{4 D_1} \right ) n^2 - \Delta t^3 \left (
\sum_{i=1}^{d} \frac{\ddot {x}_c^{i}(t)^2}{16 D_1} \right ) n^3
\right ].
\end{equation}
The integrals in Eq.(\ref{second_correction}) cannot be computed
exactly and approximations are also hard to come by, since we do not
\emph{a priori} know the magnitude of the walker's velocity $\dot
{\bf x}_c(t)$ and acceleration $\ddot {\bf x}_c(t)$, which appear in
the argument of the exponent. The main problem here is that one has
an implicit equation (given by Eq.(\ref{e1}) together with
Eq.(\ref{second_correction})) in the walker's velocity. The only
fact which can be safely deduced is that for $d \gg 2$, the two
integrals are approximately equal; this implies that the second term
in Eq.(\ref{second_correction}) vanishes in the limit $\Delta t
\rightarrow 0$ and that in high dimensions there are no further
corrections to the modified Langevin equation Eq.(\ref{mod_Lan}).
Clearly in the formalism of the Langevin equation it is not easily
possible to systematically calculate corrections to the modified
Langevin equation Eq.(\ref{mod_Lan}). However it is to be emphasized
that the derivation of the latter equation (and the subsequent
prediction of the phase transition), elude a treatment based on the
equation of motion of the walker's probability density function.

\end{document}